%% file: revol_for.tex
\documentclass[conference,letterpaper]{IEEEtran}

\usepackage{layouts}
\usepackage[T1]{fontenc}
\usepackage[utf8]{inputenc}
\usepackage[english]{babel}

\usepackage{ifthen}

\usepackage{times}
\usepackage{balance}
\usepackage{csquotes}
\usepackage{microtype}

\usepackage{xcolor}
\usepackage{graphicx}

\usepackage{bm}
\usepackage{amsmath}

\usepackage{amsthm}
\usepackage{amsfonts}
\usepackage{amssymb}

\usepackage{siunitx}
\usepackage{algorithm}
\usepackage{algpseudocode}

\usepackage[
    bibstyle=ieee,
    citestyle=numeric-comp,
    minnames=1,
    maxnames=6,
    mincitenames=1,
    maxcitenames=1,
    autocite=inline,
    sortcites=true,
    dashed=false,
    labelnumber=true]{biblatex} 
\usepackage[acronym,hyperfirst=false,nohypertypes={acronym}]{glossaries}

\usepackage[font=small,belowskip=-1.1em]{caption}

\usepackage{tabu}

\newcolumntype{C}{>{\centering\arraybackslash}X} % centered version of 'X' col. type
\usepackage{booktabs}
\usepackage{xspace}

\usepackage[disable]{todonotes}

\usepackage{url}
\usepackage[colorlinks=false,hidelinks=true]{hyperref}
\usepackage{cleveref}
\usepackage[shortcuts]{extdash}

\makeatletter
\newcounter{IEEE@bibentries}
\renewcommand\IEEEtriggeratref[1]{%
  \renewbibmacro{finentry}{%
    \stepcounter{IEEE@bibentries}%
    \ifthenelse{\equal{\value{IEEE@bibentries}}{#1}}
    {\finentry\@IEEEtriggercmd}
    {\finentry}%
  }%
}
\makeatother

\DeclareCaptionLabelSeparator{iaria}{.\quad}
\graphicspath{{./figures/}}
\DeclareGraphicsExtensions{.pdf,.jpeg,.png}

\crefname{figure}{Figure}{Figures}
\crefname{table}{Table}{Tables}
\crefname{section}{Section}{Sections}
\crefname{equation}{Equation}{Equations}
\crefname{algorithm}{Algorithm}{Algorithms}

\loadglsentries[main]{glossary_revol_for.inc.tex}

\begin{document}
% \title{\Large\bfseries%
%     Optimization-based determination of the flexible operation region of
%   a distribution grid using an evolutionary optimization heuristic}
\title{\Large\bfseries%
    Comparison of Random Sampling and Heuristic Optimization-Based Methods for 
    Determining the Flexibility Potential at Vertical System Interconnections}
%\title{\Large\bfseries%
%    Pointing out the folding problem of stochastic flexibility aggregation methods}
% \title{\Large\bfseries%
%     Pointing out the folding problem of sampling-based flexibility aggregation methods}
\author{\IEEEauthorblockN{Johannes Gerster\\ and Sebastian Lehnhoff}
  \IEEEauthorblockA{Department of Computing Science\\
    CvO Universität Oldenburg\\
  johannes.gerster@uol.de}
  \and
  \IEEEauthorblockN{Marcel Sarstedt\\ and Lutz Hofmann}
  \IEEEauthorblockA{Institute of Electric Power Systems\\
    Leibniz Universität Hannover\\
  sarstedt@ifes.uni-hannover.de}
  \and
  \IEEEauthorblockN{Eric MSP Veith}
  \IEEEauthorblockA{OFFIS e.V.\\
    R\&D Division Energy\\
    Oldenburg, Germany\\
  eric.veith@offis.de}}

\maketitle

\begin{abstract}
In order to prevent conflicting or counteracting use of flexibility options,
the coordination between distribution system operator and transmission system
operator has to be strengthened. For this purpose, methods for the standardized
description and identification of the aggregated flexibility potential of
distribution grids are developed. Approaches for identifying the \gls{FOR} of
distribution grids can be categorized into two main classes: Random
sampling/stochastic approaches and optimization-based approaches. While the
former have the advantage of working in real-world scenarios where no full grid
models exist, when relying on nai\"ve sampling strategies, they suffer from
poor coverage of the edges of the~\gls{FOR} due to convoluted
distributions. In this paper, we tackle the problem from two different sides.
First, we present a random sampling approach which mitigates the convolution
problem by drawing sample values from a multivariate Dirichlet distribution.
Second, we come up with a hybrid approach which solves the underlying optimal
power flow problems of the optimization-based approach by means of a stochastic
evolutionary optimization algorithm codenamed REvol. By means of synthetic
feeders, we compare the two proposed \gls{FOR} identification methods with
regard to how well the \gls{FOR} is covered and number of power flow
calculations required.
\end{abstract}


\section{Introduction}%
\label{sec:introduction}
% \printinunitsof{in}\prntlen{\textwidth}
The increasing share of \glspl{DER} in the electrical energy system leads to
new challenges for both, \glspl{TSO} and \glspl{DSO}. Flexible ancillary
services for congestion management, voltage maintenance or power balancing, so
far mostly provided by large scale thermal power plants directly connected to
the \gls{TG}, increasingly have to be provided by \glspl{DER} connected to the
\gls{DG}. Thus, \glspl{DG} evolve from formerly mostly passive systems to
\glspl{ADG} that contain a variety of controllable components interconnected
via communication infrastructure and whose dynamic behavior is characterized by
higher variability of vertical power flows and greater simultaneity factors.

\gls{TSO}-\gls{DSO} coordination is an important topic which has been pushed by
ENTSO\=/E during the last years~\cite{entso-e2015, entso-e2015a, entso-e2017,
entso-e2019}. Coordination between grid operators has to be strengthened to
prevent conflicting or counteracting use of flexibility
options~\cite{sarstedt2019}. To reduce complexity for \glspl{TSO} at the
\gls{TSO}/\gls{DSO} interface and to enable \glspl{TSO} to consider the
flexibility potential of \glspl{DG} in its operational management and
optimization processes, methods are needed which allow for the determination
and standardized representation of the aggregated flexibility potential of
\glspl{DG}.

The aggregated flexibility potential of a \gls{DG} can be described as region
in the PQ-plane that is made up from the set of feasible
\glspl{IPF}~\cite{mayorgagonzalez2018}. Thereby, feasible \glspl{IPF} are
\glspl{IPF} which can be realized by using the flexibilities of controllable
\glspl{DER} and controllable grid components such as \gls{OLTC} transformers in
compliance with grid constraints i.e.,\ voltage limits and maximum line
currents.

In the literature, there are various concepts to determine the \gls{FOR} of
\glspl{DG}. They can be categorized into two main classes:
Random sampling/stochastic approaches and optimization-based approaches.

For random sampling approaches in its simplest form, the general procedure is
such that a set of random control scenarios is generated by assigning
set-values from a uniform distribution to each controllable unit. By means of
load flow calculations the resulting \glspl{IPF} are determined for each
control scenario and classified into feasible \glspl{IPF} (no grid constraints
are violated) and non-feasible \glspl{IPF} (at least one grid constraint is
violated). The resulting point cloud of feasible \glspl{IPF} in the PQ-plane
serves as stencil for the \gls{FOR}~\cite{mayorgagonzalez2018}. A problem that
comes with this approach is that drawing set-values from independent uniform
distributions leads to an unfavorable distribution of the resulting \glspl{IPF}
in the PQ-plane and extreme points on the margins of the \gls{FOR} are not
captured well~\cite{gerster2021}.

This is where optimization-based methods come into play. The basic idea behind
these methods is not to randomly sample \glspl{IPF} but systematically identify
marginal \glspl{IPF} by solving a series of \gls{OPF}
problems~\cite{silva2018}. In addition to better coverage of the \gls{FOR},
optimization-based approaches have the advantage of higher computational
efficiency. An important drawback is however that, except for approaches which
solve the \gls{OPF} heuristically, solving the underlying \gls{OPF} requires an
explicit grid model~\cite{contreras2018}. On the other hand, the only heuristic
approach published so far suffers from poor automatability as it relies on
manual tweaking of hyperparameters~\cite{sarstedt2021}. 

In practice, considering the huge size of \glspl{DG}, data related to the grid
topology (asset data, operating equipment, etc.) usually does not cover the
whole grid~\cite{singh2010}, which
complicates parametrization of explicit grid models. 
In such circumstances, black-box machine learning (ML) models trained on
measurement data provided by current smart meters can be an alternative to
physics-based, explicit grid models~\cite{barbeiro2014}. 
Therefore, we argue that it is worthwhile to research \gls{FOR} 
determination techniques, which are compatible with black-box grid models.
In this paper, we tackle the problem from two different sides. 
As we have detailed the convolution problem previously~\cite{gerster2021}, we
now show a solution approach. In this vein, we provide a comparison between
a random sampling method more suited to the problem, as well as a heuristic as
informed search method.
First, we present our random sampling approach which mitigates the convolution 
problem by drawing sample values from a multivariate Dirichlet distribution.
Second, we introduce a hybrid approach which solves the underlying \gls{OPF}
problems of the optimization-based approach heuristically by means of the
evolutionary optimization algorithm REvol~\cite{veith2014}.
To show automatability, in 
contrast to~\cite{sarstedt2021}, we do no manual tweaking 
of hyperparameters. Instead, parameter optimization is done via 
random search. Furthermore, we do not make problem-specific adjustments
to the optimization algorithm.

The remainder of the paper is structured as follows: A survey on existing
approaches (random sampling and optimization-based) and the contribution of
this paper are given in \autoref{sec:related_work}. 
Next, in
\autoref{sec:benchmark_feeder}, the construction of a series of synthetic
feeders with increasing number of nodes is explained. These feeders are used for
comparison and evaluation of our proposed \gls{FOR} determination techniques. 
In the subsequent sections~\ref{sec:dirichlet-approach}
and~\ref{sec:revol-approach}, we present our stochastic
approach based on sampling from a Dirichlet distribution and our 
hybrid approach using REvol for solving the series of 
\gls{OPF} problems. In this context, we additionally 
discuss how the parameter tuning is performed and introduce the
Jaccard index as a measure for comparing \glspl{FOR}. 
After this, in \autoref{sec:results}, we present and discuss our results.
Finally, in \autoref{sec:conclusion}, the paper is summarized and the
conclusion---along with a brief outline of future work---is given.

%%%%%%%%%%%%%%%%%%%%%%%%%%%%%%%%%%%%%%%%%%%%%%%%%%%%%%%%%%%%%%%%%%%%%%%%%%%%%%%
\section{Related work and contribution of this paper}\label{sec:related_work}
As outlined in the introduction, relevant literature can be grouped into two
main categories: Random sampling/stochastic and optimization-based approaches
for exploring the \gls{FOR} of \glspl{DG}. The following two subsections are
mainly taken from~\cite{gerster2021}. In~\cite{gerster2021}, we motivate 
work presented in this paper by pointing out the
convolution problem when sampling from uniform distributions.

\subsection{Random sampling approaches}\label{sec:data_driven_approaches}
\textcite{heleno2015} are the first to come up with the idea of estimating the
flexibility range in each primary substation node to inform the \gls{TSO} about
the technically feasible aggregated flexibility of \glspl{DG}. In order to
enable the \gls{TSO} to perform a cost/benefit evaluation, they also include
the costs associated with adjusting the originally planned operating point of
flexible resources in their algorithm. In the paper two variants of a Monte
Carlo simulation approach are presented, which differ in the assignment of
set-values to the flexible resources. While in the first approach independent
random set-values for changing active and reactive power injection are
associated to each flexible resource, in the second approach a negative
correlation of one between generation and load at the same bus was considered.
In a direct comparison of the two presented approaches, the approach with
negative correlation between generation and load at the same bus performs
better and results in a wider flexibility range and lower flexibility costs
with a smaller sample size. Nevertheless, even with this approach, the
capability to find marginal points in the \gls{FOR} is limited. Therefore, in
the outlook the authors suggest the formulation of an optimization problem in
order to overcome the limitations of the Monte Carlo simulation approach,
increasing the capability to find extreme points of the \gls{FOR} and reducing
the computational effort. In \textcite{silva2018}, which is discussed in the
next subsection, the authors take up this idea again.

\textcite{mayorgagonzalez2018} extend in their paper the methodology presented
by \textcite{heleno2015}. First, they describe an approach to approximate the
\gls{FOR} of an \gls{ADG} for a particular point in time assuming that all
influencing factors are known. For this, they use the first approach of
\textcite{heleno2015} for sampling \glspl{IPF} (the one that does not consider
correlations). That is, random control scenarios are generated by assigning
set-values from independent uniform distributions to all controllable units. In
contrast to~\cite{heleno2015}, no cost values are calculated for the resulting
\glspl{IPF}. Instead, for describing the numerically computed \gls{FOR} with
sparse data, the region is approximated with a polygon in the complex plane. In
addition, a probabilistic approach to assess in advance the flexibility
associated to an \gls{ADG} that will be available in a future time interval
under consideration of forecasts which are subject to uncertainty is proposed.
The authors mention that for practical usage the computation time for both
approaches has to be significantly reduced. However, the problem of
unfavorable distribution of the resulting \gls{IPF} point cloud, when drawing
control scenarios from independent uniform distributions, which is a mayor
factor for the low computational efficiency, is not discussed.

When research presented in this paper was already advanced,
\textcite{contreras2021} came up with advanced
random sampling approaches. They show that, when focusing the
vertices of the flexibility chart of flexibility providing units during
sampling, \gls{FOR} coverage can be dramatically improved
in comparison to the nai\"ve sampling. On top of that, they present
a comparison of OPF-based and random sampling approaches, whose results show
that with their improved sampling strategies both approaches are capable of
assessing the \gls{FOR} of radial distribution grids but for grids with large
number of buses OPF-based methods are still better suited.

\subsection{Optimization-based approaches}%
\label{sec:optimization_based_approaches}
\textcite{silva2018} address the main limitation of their sampling-based
approach in~\textcite{heleno2015}, namely estimating extreme points in the
\gls{FOR}. To this end, they propose a methodology which is based on
formulating an optimization problem with below-mentioned objective function,
whose maximization for different ratios of $\alpha$ and $\beta$ allows to
capture the perimeter of the flexibility area.

\begin{equation}
    \alpha \: P_{\mathit{DSO} \rightarrow \mathit{TSO}}
    + \beta \: Q_{\mathit{DSO} \rightarrow \mathit{TSO}}
\end{equation}

\noindent where $P_{\mathit{DSO} \rightarrow \mathit{TSO}}$ and
$Q_{\mathit{DSO} \rightarrow \mathit{TSO}}$ are the active and reactive power
injections at the TSO-DSO boundary nodes. \textcite{silva2018} work out that
the underlying optimization problem represents an OPF problem. Due to its
robust characteristics they use the primal-dual, a variant of the interior
point methods to solve it.
The methodology was evaluated in simulation and validated in real field-tests
on MV distribution networks in France. The comparison of simulation results
with the random sampling algorithm in~\textcite{heleno2015} shows the
superiority of the optimization-based approach by illustrating its capability
to identify a larger flexibility area and to do it within a shorter computing
time.

\textcite{capitanescu2018} propose the concept of active-reactive power (PQ)
chart, which characterizes the short-term flexibility capability of active
distribution networks to provide ancillary services to \gls{TSO}. To support
this concept, an AC optimal power flow-based methodology to generate PQ
capability charts of desired granularity is proposed and illustrated in
a modified 34-bus distribution grid.

\textcite{contreras2018} present a linear optimization model for the
aggregation of active and reactive power flexibility of distribution grids at
a \gls{TSO}-\gls{DSO} interconnection point. The power flow equations are
linearized by using the Jacobian matrix of the Newton-Raphson algorithm. The
model is complemented with non-rectangular linear representations of typical
flexibility providing units, increasing the accuracy of the distribution grid
aggregation. The obtained linear programming system allows a considerable
reduction of the required computing time for the process. At the same time, it
maintains the accuracy of the power flow calculations and increases the
stability of the search algorithm while considering large grid models.

\textcite{fortenbacher2020} present a method to compute reduced and aggregated
distribution grid representations that provide an interface in the form of
active and reactive power (PQ) capability areas to improve \gls{TSO}-\gls{DSO}
interactions. Based on a lossless linear power flow approximation they derive
polyhedral sets to determine a reduced PQ operating region capturing all
voltage magnitude and branch power flow constraints of the \gls{DG}. While
approximation errors are reasonable, especially for low voltage grids,
computational complexity is significantly reduced with this method.

\textcite{sarstedt2021} provide a detailed survey on stochastic and
optimization based methods for the determination of the \gls{FOR}. They come up
with a comparison of different \gls{FOR} determination methods with regard to
quality of results and computation time. For their comparison they use the
Cigr\'e medium voltage test system. On top of that, they present a \gls{PSO}
based method for \gls{FOR} determination. 

\subsection*{Contributions of this paper}
In summary, it can be stated that optimization-based approaches show high
computational efficiency with good coverage of the \gls{FOR}. However, methods
used for solving the underlying \gls{OPF} problem rely---except for heuristic
approaches---on explicit grid models of the
\gls{DG}, which must be parametrized with grid topology data often not
available in practice. At the same time, the only heuristic approach
presented so far, suffers from poor automatability due to manual
tweaking of hyperparameters~\cite{sarstedt2021}.
Random sampling approaches, on the
other hand, do not require explicit grid modeling and are compatible with
black-box grid models, but suffer from low computational efficiency and poor
coverage of peripheral regions of the \gls{FOR}, when using conventional
sampling strategies.

In this paper, we tackle the problem from two different sides.
First, we present a random sampling approach which mitigates the convolution 
problem. In a two-stage procedure, we start with drawing set-values
from a uniform distribution
for the aggregated active and reactive power of the flexible units 
(inverter-connected batteries in our case). In a second step, 
the previously determined set-values for the aggregated power are
distributed to the individual units. For this, the Dirichlet distribution
is used. Here, we make use of the property that the sum of the 
entries of the random
vectors generated from the multivariate Dirichlet distribution is
always one.

Our second approach is to make the optimization-based
approach black-box model compatible
by solving the underlying \gls{OPF} problem heuristically with the
evolutionary optimization algorithm REvol~\cite{veith2014}.
To show automatability,
we use randomized search for hyperparameter tuning.
As score for parameter optimization we take the Jaccard index
to measure similarity between \glspl{FOR} determined by REvol with 
varying parameter combinations and a benchmark \gls{FOR}, which 
is generated by applying the Dirichlet sampling. To get close to the
actual \gls{FOR}, a large sample size (\num{100000} sample elements)
is used for creating the benchmark \gls{FOR}.

%%%%%%%%%%%%%%%%%%%%%%%%%%%%%%%%%%%%%%%%%%%%%%%%%%%%%%%%%%%%%%%%%%%%%%%%%%%%%%%
\section{Synthetic benchmark feeders and
comparison scenario}%
\begin{figure}
  \includegraphics[width=\linewidth]{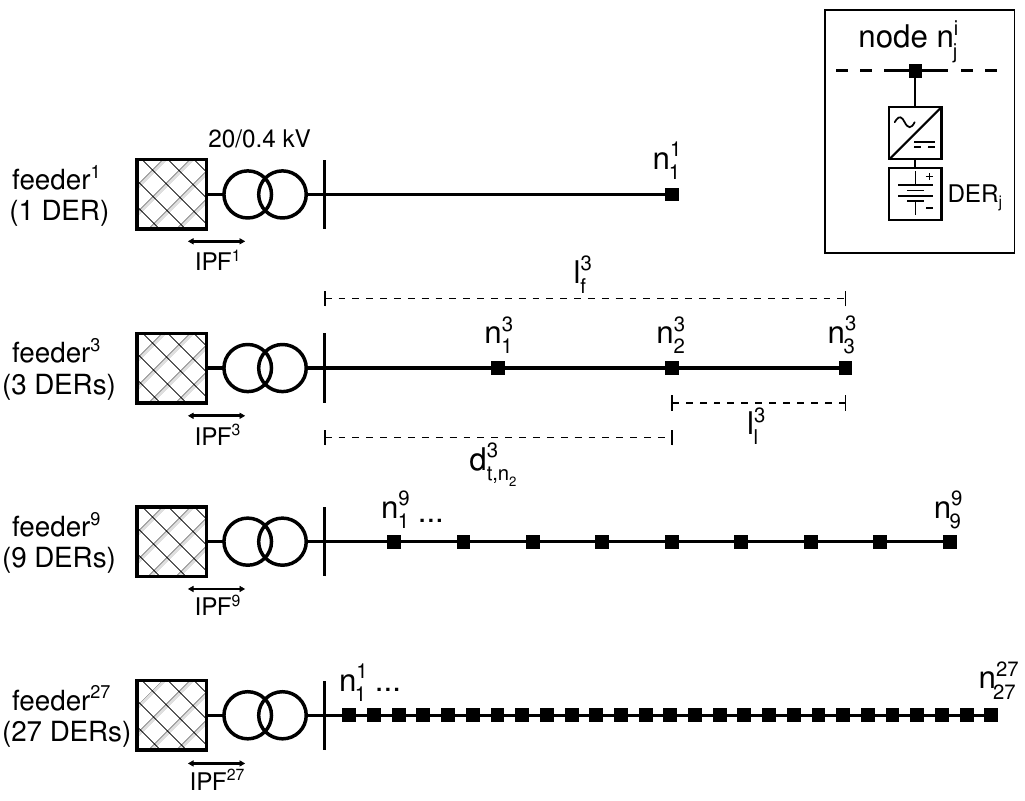}
  \caption{Synthetic \SI{0.4}{kV} feeders}
  \label{fig:grid_sketch}
\end{figure}

\label{sec:benchmark_feeder}
Evaluation and comparison of the \gls{FOR} identification
methods presented in this paper are done on the basis of
a series of synthetic \SI{0.4}{kV} feeders as shown in
\cref{fig:grid_sketch}. Construction of the synthetic feeders
is explained in more detail in~\cite{gerster2021}. 
The feeders are characterized by an increasing number
of nodes. In order to better evaluate how an increasing number of
nodes affects the investigated \gls{FOR} determination methods,
both, the total installed power and
the mean transformer-node distance are chosen equal for all feeders.
This results in similar, easily comparable \glspl{FOR}
regardless of the number of
nodes.

For our experiments, we have constructed four synthetic \SI{0.4}{kV}~feeders.
The feeders differ in the number of nodes $N^i$, which has been set to \num{1},
\num{3}, \num{9} or \num{27} respectively.
To get similar \glspl{FOR} for all feeders, 
line length $l^i_l$ and feeder
length $l^i_f$ have then been calculated in such a way, that the 
mean transformer-node distance is equal for all feeders. The resulting
line parameters
are shown in \cref{tab:grids_configuration}. Formulas for 
calculating the line lengths, can be found in~\cite{gerster2021}. 
There is one \gls{DER} connected to each node and the
installed power $P^i_{inst, DERs}$ is distributed
evenly among the \glspl{DER}. To be able to cover the entire
flexibility area of the feeders including their border areas where voltage band
violations and/or line overloadings can be observed, all \glspl{DER} are
inverter-connected battery storages because they offer maximum flexibility with
regard to both, active and reactive power provision.
The dimensioning of the
inverters has been chosen in such a way that a power factor $\cos{\phi}$ of
0.9~inductive can be kept, when the maximum active power is delivered: Values for
the technical parameters of the four feeders including connected \glspl{DER}
are listed in \cref{tab:grids_configuration}. 

\begin{table*}[t]
\caption{Configuration of the synthetic feeders}
\label{tab:grids_configuration}
  \begin{tabu}{CCCCCCCC}
    \toprule
    \# DERs & $\mathrm{P_{inst, DER_j}}$ (kW) & $\mathrm{|S|_{max,DER_j}}$ (kVA)
    & Feeder Length (m) & Line Length (m)
    & Line Type & Voltage Band (pu) & Trafo Type\\
    \midrule
    1 & 200.0 & 222.2 & 400 & 400 & NAYY 4x150 SE & 
    0.9--1.1 & 0.4 MVA 20/0.4~kV \\
    3 & 66.7  & 74.1  & 600 & 200 & NAYY 4x150 SE & 
    0.9--1.1 & 0.4 MVA 20/0.4~kV \\
    9 & 22.2  & 24.7  & 720 & 80  & NAYY 4x150 SE & 
    0.9--1.1 & 0.4 MVA 20/0.4~kV \\
    27 & 7.4  & 8.2   & 771 & 29  & NAYY 4x150 SE & 
    0.9--1.1 & 0.4 MVA 20/0.4~kV \\
    \bottomrule
  \end{tabu}
\end{table*}

%%%%%%%%%%%%%%%%%%%%%%%%%%%%%%%%%%%%%%%%%%%%%%%%%%%%%%%%%%%%%%%%%%%%%%%%%%%%%%%
\section{Two-stage random sampling approach with sampling from Dirichlet
distribution}%
\label{sec:dirichlet-approach}
To mitigate the problem of convoluted distributions when using
classical sampling strategies, as it was pointed out in~\cite{gerster2021},
we propose a two-stage procedure. It is motivated by the following 
basic consideration: During sampling, the \gls{FOR} should be covered 
evenly. As \glspl{IPF} are mostly made
up of the sum of power injections of connected \glspl{DER} (grid
losses are comparatively small), it is reasonable to draw set-values
for the aggregated power of connected \glspl{DER} in a first step
and then distribute them randomly to the individual units. 
To do so, we need an appropriate distribution which allows for 
sampling random vectors with given sum (to match the set value
for the aggregated power drawn in the first step). As it has
exactly this property, this is where
the Dirichlet distribution $\operatorname{Dir}(\boldsymbol\alpha)$
comes into play. It is a continuous multivariate
probability distribution parameterized by a
vector~$\boldsymbol\alpha$ of positive reals, which controls the
concentration of the distribution:

\begin{equation}
  \label{eq:dirichlet_pdf}
 \operatorname{Dir}(\boldsymbol\alpha)
   = f \left(x_1,\ldots, x_{K}; \alpha_1,\ldots, \alpha_K \right)
\end{equation} 

\noindent
where:
\begin{equation}
  \label{eq:dirichlet_sum}
  \sum_{i=1}^{K} x_i = 1 \mbox{ and } x_i \ge 0 \mbox{ for all } i \in \{1,\dots,K\}
\end{equation} 
\noindent
with $K$ indicating the order of the multivariate distribution.
For the details regarding the Dirichlet distribution, please refer to
statistics text books e.g.,~\cite{ng2011dirichlet}.

The power set-values for the individual units generated by means of the
Dirichlet distribution are not distributed uniformly. The shape of the
distribution can be manipulated via the~$\boldsymbol\alpha$~parameter.
With $\alpha_i = 1$ for all $i \in \{1,\dots,K\}$, the distribution of 
the individual set-values is biased towards low power values. 
During an initial experimenting phase, by trial and error,
$\alpha_i = 1.2$ for all $i \in \{1,\dots,K\}$ has turned out to
be good. Systematic optimization of
$\boldsymbol\alpha$ is part of future work.

There is another factor biasing the distribution of \glspl{IPF}. 
Due to the fact, that the random power set-values for the 
individual units generated by means of the
Dirichlet distribution are not all in the power range that can be 
realized by the individual units and set-values beyond the limits
are clipped by our simulations. This would result in an accumulation of
\glspl{IPF} at the upper edge of the \gls{FOR}. 

To reduce the aforementioned distortions, before applying the normalized
set-values to our simulation environment, we split the sample in four subsets
of equal size. The first subset remains unchanged, for the second subset, 
normalized active power values $\boldsymbol{p_n}$
are transformed with $\boldsymbol{p_n'=-p_n' + 1}$, for the
third subset, the transformation is done to the reactive power set-values 
$\boldsymbol{q_n}$ ($\boldsymbol{q_n'=-q_n + 1}$).
Finally, with the fourth subset, the transformation
is applied to both, active and reactive power set-values.

We proceed in 
such a way, that we calculate the absolute active 
and reactive power set-values from the normalized ones generated
during sampling. After assigning these values to 
the \glspl{DER}, we step the rudimentary inverter simulators
via which all batteries are connected to the synthetic feeders.
The inverter models monitor that the maximum current of the
inverters is not exceeded.
In case the provided set point is not compatible with
the current limits of the inverter, it is operated at 
an operating point which is as close as possible to the preset, 
but still complies with the current constraint. The power 
values returned from the inverter simulators are than passed 
to the \emph{pandapower} library~\cite{thurner2018a} which is used for
calculating the power flows.  This way, we generate a sample of size
\num{10000} for each feeder.

Following this, the sample elements are first classified with regard to their
adherence to grid constraints and in case of non-adherence with regard to the
type of grid constraint violation (i.e., voltage band violation, line overload,
or both). In the final step, we compute the convex hull around the
\glspl{IPF} which do not violate any constraints and thus make up the
\gls{FOR}.

\begin{figure}
\centering
  \includegraphics{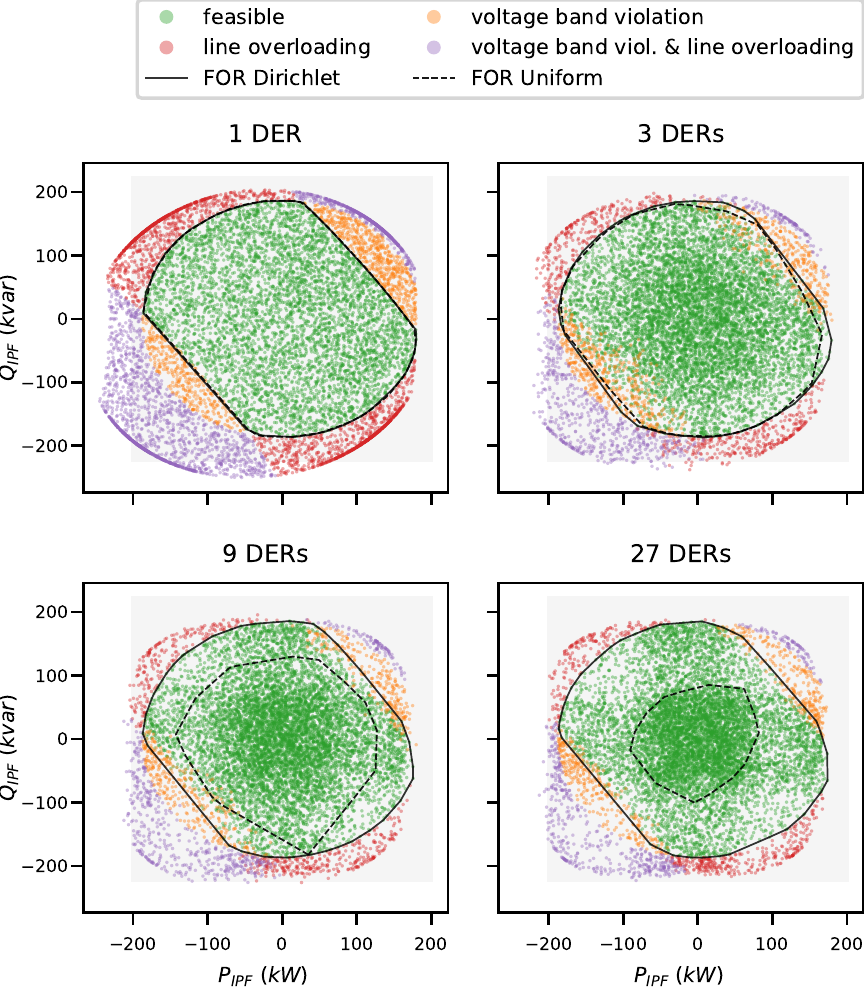}
  \caption{Results of sampling from Dirichlet distribution (\num{10000} sample
    elements) classified by feasibility with regard to grid constraints
    (voltage band limits and max.\ line loading); black solid line indicates
    the resulting \gls{FOR}; black dotted line indicates \gls{FOR} when
    sampling from uniform distribution with same sample size}%
  \label{fig:eval_dirichlet}
\end{figure}

The resulting \glspl{FOR} for feeders with increasing number of 
nodes are shown in \cref{fig:eval_dirichlet}. The border of the \gls{FOR} is
marked by the solid black line in each case. For comparison, the dashed lines 
define the \glspl{FOR} which result from sampling from a uniform distribution.
It can be seen from the plots that with the
Dirichlet approach the identified \gls{FOR} is approximately the same for 
all tested feeders. A collapsing \gls{FOR}, as observed with the uniform
sampling approach (dotted lines), does not occur.
This indicates that the folding problem, we pointed out in~\cite{gerster2021},
is mitigated with the procedure proposed here. 

Nevertheless, it can be easily
seen, that even with the Dirichlet approach, the \gls{FOR} is not evenly
covered. As the number of nodes increases, clusters are formed around the
coordinate origin and along the axes.  This is particularly problematic in that
fewer points are sampled in the interesting areas in the first and third
quadrant, where due to voltage band violations the density of valid \glspl{IPF}
decreases anyway. The reason probably lies in the transformation of the data
along the main axes, which is motivated by mitigating the bias caused by
clipping of set-values beyond the power limits of the \glspl{DER}. A~more
suitable data transformation and an optimized selection of the~$\boldsymbol
\alpha$~parameter of the Dirichlet distribution could remedy this and offer
potential for future work.

We use the two-stage Dirichlet approach presented in this section
to generate the benchmark \gls{FOR} for the parameter tuning of the
optimization-based approach which will be introduced in the next 
section. In contrast to the results presented here, to get as close as possible
to the actual \gls{FOR}, we increase the 
sample size from \num{10000} to \num{100000}.

%%%%%%%%%%%%%%%%%%%%%%%%%%%%%%%%%%%%%%%%%%%%%%%%%%%%%%%%%%%%%%%%%%%%%%%%%%%%%%%
\section{Optimization-based approach with multipart evolutionary algorithm REvol}
\label{sec:revol-approach}
Metaheuristics are a common tool in the field of electric power system
optimization and have been used to solve various optimization problems regarding
system planning and operation (see \cite{wgmho2021}). In \cite{sarstedt2021}
for the first time a metaheuristic was used in the context of
\gls{FOR} determination.
Due to its popularity in solving \gls{OPF} problems, \gls{PSO} was 
applied.
For improved performance, in \cite{sarstedt2021}
the \gls{PSO} is tailored to the \gls{FOR} identification
problem and hyperparameters are tweaked manually. Because automation
suffers from that, in this paper we choose a different approach and 
dispense with problem specific customization of the applied metaheuristic
and do automated hyperparameter tuning by means of the randomized search
facility provided by the Python machine learning framework
Scikit-learn \cite{sklearn2021}.
Due to its superior performance compared to \gls{PSO} on
other tasks (training of \glspl{ANN} for time series prediction),
in this paper
we give REvol a try in optimization based \gls{FOR} identification.
REvol was introduced in~\cite{veith2014}. 
Comparison with \gls{PSO} regarding training of \glspl{ANN} can be
found in~\cite{veith2016}. The following statements are 
taken in abbreviated form from~\cite{veith2016} and can be read there
in detail.

\subsection{REvol basics} 
REvol employs an approach similar to the classic
evolutionary algorithms in that it defines a number of individuals that live
in a population. In contrast to the classical approach, an individual 
consists of two vectors: A parameter and a scatter vector, each with the
same size. While the parameter vector represents a solution candidate 
for the problem 
at hand, in our case normalized active and reactive power set-values for 
each battery connected to the feeder, each component of the scatter vector
limits the variability of the corresponding parameter vector's component.
Also, the evolution process is basically the same as for other evolutionary 
algorithms: Until the number of iterations reaches a predefined maximum,
the algorithm generates a new individual, uses a fitness function to
evaluate it, and possibly enhances the population with it. 
However, when it comes to the creation of new individuals
REvol has two distinctive features that set it apart from the classic
approach:
 
\begin{itemize}
    \item The current rate of success
    \item The implicit gradient information
\end{itemize}
The current rate of success influences the spread, i.e., the area within which
a new individual can potentially be placed. By increasing the spread when the
current notion of success is greater then the target success, this helps to
escape local minima. In order to avoid fluctuations in the success variable, 
it is dynamically averaged over a fixed number of iterations, represented by
the user supplied variable T. The averaging is done by a time-discrete PT1
element: 

\begin{equation}
 \label{eq:pt1_element}
 pt1(y,u,t) = 
\begin{cases}
  u & \text{if } t=0   \\
  y + \frac{u-y}{t} & \text{otherwise}
\end{cases}
\end{equation} 
\noindent with $t=T$, except for the initial iteration with $t=0$. 

The second factor influencing reproduction, is the gradient
information. The area within which new individuals can be placed does not
feature a uniform \gls{PDF}. Instead, the placement of the two individuals
chosen relative to each other is used to calculate the implicit gradient, i.e.,
the direction within which a newly generated individual must be placed with
a high probability in order to reach the minimum. These two factors carefully
balance each other: Putting a strong emphasis on the implicit gradient
information would turn the multi-part evolutionary strategy into a `poor man’s
gradient decent, whereas a high influence of the dynamic reproduction
probability density spread will make the algorithm lose its orientation. This
is for providing a basic understanding of the algorithm. For the technical
details, please refer to~\cite{veith2014} and~\cite{veith2016}.

\subsection{Sampling strategy and optimization problem}
Optimization-based methods for determining the \gls{FOR} are based on
systematically identifying boundary points of the \gls{FOR}, where the 
identification of each boundary point represents an optimization problem.
At this point, it is important to emphasize that optimization-based
approaches are characterized by two important features. First, the
systematics which is used to determine the boundary points. In literature, this
is also referred to as sampling strategy~\cite{sarstedt2021}. The sampling
strategy specifies the rules by means of which a series of optimization
problems is constructed for systematically determining the \gls{FOR}. The
other important feature characterizing optimization-based \gls{FOR}
identification techniques, is how the resulting series of optimization problems
is actually solved, i.e.~which optimization method is applied. 
In this paper, the focus is on the latter. Therefore, a simple sampling
strategy is applied, whose basic function is explained in the next
paragraph. For a detailed overview of sampling strategies, please refer
to~\cite{sarstedt2021}.

The sampling strategy used for the experiments in this paper is based on 
solving the optimization problem in \cref{eq:opt_problem} 
for different ratios of $\alpha$ and $\beta$. 

\begin{equation}\label{eq:opt_problem}
  \max (\alpha \: P_{\mathit{DSO} \rightarrow \mathit{TSO}}
  + \beta \: Q_{\mathit{DSO} \rightarrow \mathit{TSO}})
\end{equation}

\noindent
where $P_{\mathit{DSO} \rightarrow \mathit{TSO}}$ and
$Q_{\mathit{DSO} \rightarrow \mathit{TSO}}$ are the active and reactive power
injections at the TSO-DSO boundary nodes. Decision variables are the active and
reactive power set points for the connected \glspl{DER} $P_{set,DER_j}$ and
$Q_{set,DER_j}$. The optimization problem is subject to compliance with grid
constraints, which are defined by the minimum and maximum voltage limits, the
maximum thermal current limits of the lines and the rated load of the
transformer. For our experiments, we determine eight boundary points with:

% \begin{equation}\label{eq:alpha_beta_cominations}
\begin{align}
  \begin{split}
    (\alpha, \beta) \in [&(1, 0), (1,1), (0,1), (-1,1), \\
                         &(-1,0), (-1,-1), (-1,0), (1,-1)] 
  \end{split}
\end{align}
% \end{equation}

\subsection{Optimization process and restrictions vector}
When evaluating individuals during the optimization process, we proceed in the
same way as we do with the random sampling based approach in the previous
section. However, this time normalized active and reactive power set-values
are provided in form of the individual's parameter vector to the simulation.
Again, before calculating the power flow with the 
\emph{pandapower} library~\cite{thurner2018a}, the inverter simulation model ensures
adherence to inverter constraints and updates the parameter vector accordingly.
Here, it is important to mention that besides the restrictions vector, which
will be introduced in the next paragraph, the updated parameter vector is
returned to REvol. It is therefore ensured that all individuals comply with
inverter constraints and thus inverter constraint are omitted in the
restrictions vector.

\begin{algorithm}
  \begin{algorithmic}%
    \caption{Comparison operator}%
    \label{alg:comparison_operator}
    \Procedure{isBetterThan}{$indiv0, indiv1$}
      \For{$i\gets 1$ to $indiv0.restr.length$}
        \If{$(indiv0.restr[i] < indiv1.restr[i])$}
          \State \Return $true$
        \ElsIf{$(indiv0.restr[i] > indiv1.restr[i])$}
          \State \Return $false$
        \EndIf
        \EndFor
      \If{$(indiv0.restr[0] > indiv1.restr[i])$}
        \State \Return $true$
      \Else
        \State \Return $false$
      \EndIf
    \EndProcedure
  \end{algorithmic}
\end{algorithm}

REvol uses the restrictions vector for comparison of individuals. Comparison 
is done when calculating the current rate of success mentioned
before and when determining the elite. The elite is composed of the fittest
individuals of the population. It plays a role when selecting the parent
individuals for the next generation---one individual from the elite and one
from the general population are chosen as parent individuals by the algorithm.
The restrictions vector has three entries: The first entry holds the
individuals fitness, calculated according to \cref{eq:opt_problem}. In 
further entries, grid constraint violations are considered---the second entry
holds the largest voltage band violation, the third entry the highest
exceedance of the line current limits.
The comparison operator \textsc{isBetterThan}$(indiv0,indiv1)$ is shown in
\cref{alg:comparison_operator}. 
From the pseudocode, it can be seen that small or no grid constraint
violations have the highest priority during comparison. Only if both
individuals do not have any or the same grid constraint violations, the
individual with the higher fitness value is evaluated as better. 

\subsection{Hyperparameter search}
Like most metaheuristics, REvol has a number of hyperparameters. 
They are listed in \cref{tab:revol_params}.
To find suitable values for these parameters, parameter search is
performed with the randomized search facility provided by the Python machine
learning framework Scikit-learn~\cite{sklearn2021}. For comparing the quality
of different parameter combinations, an appropriate metric or score is
required which indicates how well the actual \gls{FOR} is reassembled
depending on the parameter selection. We use the Jaccard index for this
purpose. The Jaccard index is a statistic for measuring the similarity of
sample sets and is defined as the size of the intersection divided by the size
of the union of the sample sets:
\begin{equation}
  \label{eq:jaccard_index}
  J(A,B) = {{|A \cap B|}\over{|A \cup B|}}.
\end{equation} 
The reference \gls{FOR} is determined according to the two-stage Dirichlet
approach presented in \cref{sec:dirichlet-approach}. To get close to the actual
\gls{FOR}, a large sample size of \num{100000} is chosen for that purpose. It
is important to mention, that the parameter search is not done for each
individual feeder presented in \cref{sec:benchmark_feeder}. Instead, we limit
the parameter search to the \num{9} node feeder and assume that the REvol
parameters found for this feeder are also suitable for other networks.
A total of \num{210} parameter settings are sampled from uniform distributions.
To account for the stochastic nature of the algorithm, three runs are
performed for each parameter selection. The range from which the individual
parameters are sampled and the best parameter combination detected, can be
taken from \cref{tab:revol_params}. For the best parameter combination, the
Jaccard index results in a mean value of \num{0.923} with a standard
deviation of \num{0.0061}. An exact coverage of the benchmark \gls{FOR} would
result in a Jaccard index of \num{1.0}. Since we determine only eight
boundary points, this value cannot be reached, of course. 

% \begin{table}[t]
\begin{table}[t]
\caption{REvol parameter selection}%
\label{tab:revol_params}
  \begin{tabu}{CCC}
    \toprule
    Parameter & Range & Best \\
%                       & Default \\ 
    \midrule
    population size                & \num{20}--\num{40}        & \num{37}    \\
%                                                               & \num{30}    \\
    elite size                     & \num{2}--\num{5}          & \num{3}     \\
%                                                               & \num{3}     \\
    max.\ epochs                   & \num{500}--\num{20000}    & \num{16245} \\
%                                                               & \num{10000} \\
    max.\ no success epochs        & \num{20}--\num{20000}     & \num{9281}  \\ 
%                                                               & \num{2000}  \\
    T                              & \num{10}--\num{20000}     & \num{5338}  \\
%                                                               & \num{1000}  \\
    start time to live             & \num{80}--\num{20000}     & \num{763}   \\
%                                                                & \num{150}   \\
    gradient weight                & \num{0.0}--\num{3.0}      & \num{2.87}  \\
%                                                               & \num{1.00}  \\
    success weight                 & \num{0.0}--\num{3.0}      & \num{2.18}  \\
%                                                               & \num{1.00}  \\
    target success                 & \num{0.1}--\num{0.4}      & \num{0.29}  \\
%                                                               & \num{0.25}  \\
    max.\ scatter  relative        & \num{0.2}--\num{3.0}      & \num{1.74}  \\
%                                                               & \num{1.00}  \\

    \bottomrule
  \end{tabu}
\end{table}

%%%%%%%%%%%%%%%%%%%%%%%%%%%%%%%%%%%%%%%%%%%%%%%%%%%%%%%%%%%%%%%%%%%%%%%%%%%%%%
\section{Evaluation and comparison of proposed FOR identification 
techniques}%
\label{sec:results}

\begin{figure}
  \centering
  \includegraphics{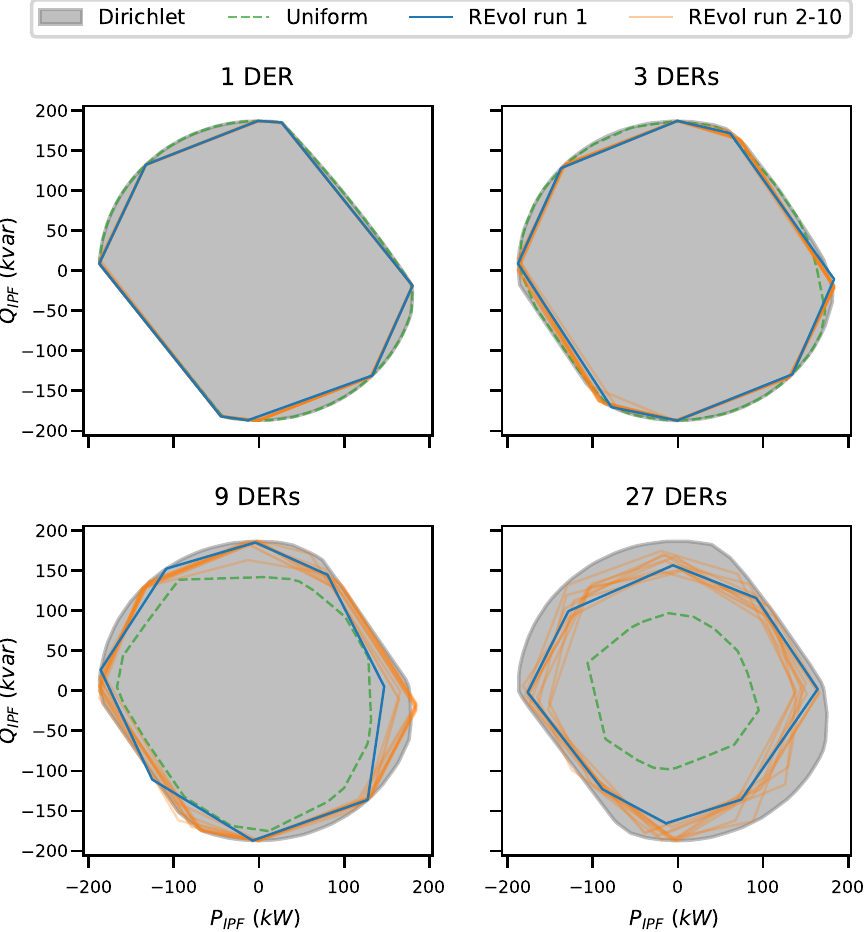}
  \caption{Comparison of \glspl{FOR} resulting from different \gls{FOR}
    identification techniques (random sampling Dirichlet, random sampling
    Uniform and optimization-based REvol) for feeders with increasing number of
    nodes}%
    \label{fig:comparison_dirichlet_revol}
\end{figure}

The plots in \cref{fig:comparison_dirichlet_revol} show the 
\glspl{FOR} resulting from the random sampling Dirichlet and the
optimization-based REvol approach in comparison. For a better understanding of
the results, the plots also include the \glspl{FOR} that can be obtained 
when sampling from uniform distributions. The comparison was
performed on the basis of the synthetic feeders described in
\cref{sec:benchmark_feeder}. The benchmark feeders are characterized by an
increasing number of nodes and connected \glspl{DER}. This allows to
investigate how well the proposed \gls{FOR} identification techniques cope with
increasing dimensionality of the search space. For the Dirichlet approach and
when sampling from uniform distributions, the
sample size was set to \num{100000}. The REvol parameters are allocated with
the best parameters found during parameter search. The exact values can be
taken from the third column in \cref{tab:revol_params}. To account for the
stochastic nature of REvol and to be able to evaluate the reliability of the
procedure, \num{10} runs have been performed with the REvol approach. Because
the borders of the \glspl{FOR} resulting from different runs are difficult to
distinguish in the plots, the first run is highlighted with its own color (blue
solid line). The two upper plots in \cref{fig:comparison_dirichlet_revol} show
that for feeders with small number of \glspl{DER}, the REvol approach delivers
good and reliable results. Apart from few exceptions in the \num{3} \glspl{DER}
case, the detected boundary points lie on the border of the \gls{FOR}
determined with the Dirichlet approach or even slightly outside, which shows
that the Dirichlet approach does not completely capture the actual \gls{FOR}
either.  Nevertheless, parts of the \gls{FOR} are not covered with the REvol
based approach. But this is mainly due to the basic sampling method and the
limitation to eight boundary points. From the two plots below, it can be seen
that the solution quality with REvol is still decent in comparison to the
random sampling from uniform distributions but decreases with
increasing number of nodes. Especially with the \num{27} nodes feeder, some
boundary points of some runs are well within the \gls{FOR} determined with the
two-stage Dirichlet random sampling approach. Moreover, from the increasing
spread of the orange lines in \cref{fig:comparison_dirichlet_revol} it can be
taken that the variance of the identified \glspl{FOR} increases with increasing
number of nodes.

The runtime of both, the Dirichlet and the REvol approach is dominated by the
power flow calculations which are performed when evaluating sample elements
and fitness function respectively. Thus, the number of load flow calculations
is a good indicator for runtime comparison. The Dirichlet approach also
performs better in this respect. The evaluation of the REvol optimization
processes for each boundary point in each run and each feeder has revealed that
in each case the maximum number of \num{16245} epochs were cycled through. With
eight boundary points, this results in about \num{130000} power flow
calculations compared to \num{100000} power flow calculations with the
two-stage Dirichlet approach and when sampling from uniform distributions.

%%%%%%%%%%%%%%%%%%%%%%%%%%%%%%%%%%%%%%%%%%%%%%%%%%%%%%%%%%%%%%%%%%%%%%%%%%%%%%%
\section{Conclusion and Future Work}%
\label{sec:conclusion}
Aggregating the flexibility potential of \glspl{DG} is an important
prerequisite for effective \gls{TSO}-\gls{DSO} coordination in electric power
systems with high share of generation located in the \gls{DG} level. In this
paper, we first gave an overview of existing flexibility aggregation methods and
categorized them in terms of whether they are random sampling/stochastic or
optimization-based. Following this, we discussed the strengths and weaknesses
of both approaches (stochastic and optimization-based) and motivated the
investigation of black-box grid model compatible \gls{FOR}~determination
techniques.
Towards this, we presented an improved random sampling based approach which
mitigates the convolution problem by drawing sample values from a multivariate
Dirichlet distribution. Second, we made the optimization-based approach
black-box compatible by applying the heuristic evolutionary algorithm REVol for
solving the underlying \gls{OPF} problem. To show automatability, we used
randomized search for hyperparameter selection.

Comparison of the two presented approaches has shown that both provide decent
coverage of the \gls{FOR}, whereby the random sampling Dirichlet approach
produces overall better results at lower runtimes.
With the REvol approach, however, we still see potential for improvement with
regard to the runtime by considering the runtime as a second factor during
parameter search besides \gls{FOR} coverage. 
Furthermore, in future work we want to investigate if the total number of
required fitness function evaluations can be reduced when sampling the border
of the FOR in one run by dynamically adapting the underlying fitness function.
In addition to improving the runtime of the REvol approach, in future work we
will compare it with other metaheuristics such as \gls{PSO}
and \gls{CMA-ES} on a more realistic benchmark grid.

% Optimierung der resultierenden Verteilung, Ripleys K als Metrik 
\section*{Acknowledgements} This work was funded by the Deutsche
Forschungsgemeinschaft (DFG, German Research Foundation) -- 359921210.

\balance

%\renewcommand*{\bibfont}{\small}
%\IEEEtriggeratref{28}
\printbibliography
\end{document}

%% file: glossary_revol_for.inc.tex
\newacronym[shortplural={CPS}]{CPS}{CPS}{cyber-physical system}
\newacronym{AI}{AI}{Artificial Intelligence}
\newacronym{AL}{AL}{Adversarial Learning}
\newacronym{ANN}{ANN}{Artificial Neural Network}
\newacronym{ARL}{ARL}{Adversarial Resilience Learning}
\newacronym{GAN}{GAN}{Generative Adversarial Network}
\newacronym{GPGPU}{GPGPU}{General-Purpose Graphics Processing Unit}
\newacronym{DL}{DL}{Deep Learning}
\newacronym{GRU}{GRU}{Gated Recurrent Unit}
\newacronym{ICT}{ICT}{Information and Communication Technology}
\newacronym{LSTM}{LSTM}{Long-Short Term Memory}
\newacronym{AEG}{AEG}{Attack Execution Graph}
\newacronym{ML}{ML}{Machine Learning}
\newacronym{PoC}{PoC}{Proof-of-Concept}
\newacronym{RL}{RL}{Reinforcement Learning}
\newacronym{DNN}{DNN}{Deep Neural Network}
\newacronym{RNN}{RNN}{Recurrent Neural Network}
\newacronym{CNN}{CNN}{Convolutional Neural Network}
\newacronym{SIMD}{SIMD}{Single Instruction, Multiple Data}
\newacronym{DoE}{DoE}{design of experiments}
\newacronym{DNC}{DNC}{Differentiable Neural Computer}
\newacronym{PV}{PV}{Photovoltaic}
\newacronym{MV}{MV}{medium voltage}
\newacronym{CHP}{CHP}{combined heat and power}
\newacronym{STL}{STL}{Signal Temporal Logic}
\newacronym{MTL}{MTL}{Metric Temporal Logic}
\newacronym{MITL}{MITL}{Metric Interval Temporal Logic}
\newacronym{AV}{AV}{Autonomous Vehicle}
\newacronym[shortplural={MAS}]{MAS}{MAS}{Multi Agent System}
\newacronym{RTU}{RTU}{Remote Terminal Unit}
\newacronym{TCP}{TCP}{Transmission Control Protocol}
\newacronym{LPEP}{LPEP}{Lightweight Power Exchange Protocol}
\newacronym{FMI}{FMI}{Functional Mock-Up Interface}
\newacronym{XML}{XML}{eXtensible Markup Language}
\newacronym{IoT}{IoT}{Internet of Things}
\newacronym{SIEM}{SIEM}{Security Information and Event Management}
\newacronym{APT}{APT}{Advanced Persistent Threads}
\newacronym{TAL}{TAL}{Thread Agents Library}
\newacronym{CBN}{CBN}{Causal Bayesian Network}
\newacronym{DER}{DER}{distributed energy resource}
\newacronym{CIGRE}{CIGRE}{International Council on Large Electric Systems}
\newacronym{IPF}{IPF}{interconnection power flow}
\newacronym{PDF}{PDF}{probability density function}
\newacronym{TSO}{TSO}{transmission system operator}
\newacronym{DSO}{DSO}{distribution system operator}
\newacronym{TG}{TG}{transmission grid}
\newacronym{DG}{DG}{distribution grid}
\newacronym{ADG}{ADG}{active distribution grid}
\newacronym{OLTC}{OLTC}{on-load tap changer}
\newacronym{FOR}{FOR}{feasible operation region}
\newacronym{OPF}{OPF}{optimal power flow}
\newacronym{PSO}{PSO}{particle swarm optimization}
\newacronym{COHDA}{COHDA}{Combinatorial Optimization
Heuristic for Distributed Agents}
\newacronym{CMA-ES}{CMA-ES}{Covariance Matrix Adaption Evolution Strategy}

%\newacronym{PYRATE}{PYRATE}{Polymorphe Agenten als querschnittliche Softwaretechnologie zur Analyse der Betriebssicherheit von cyber-physischen Systemen}
\newacronym{PYRATE}{PYRATE}{Polymorphic agents as cross-sectional software technology for the analysis of the operational safety of cyber-physical systems}

\newglossaryentry{resilience}{%
  name=resilience,
  description={The ability of a system to withstand unforseen, rare and
  potentially catastrophic events and to self-heal or improve in
  reaction to these events. Ideally resilience is increasing monotonously.}
}

% vim:ft=tex